\documentclass[11pt,fleqn]{iopart}

\usepackage{iopams}
\usepackage{fouriernc}
\usepackage{bm}

\expandafter\let\csname equation*\endcsname\relax

\expandafter\let\csname endequation*\endcsname\relax

\usepackage{amsmath}
\usepackage{amsthm}

\theoremstyle{remark}
\newtheorem{Remark}{Remark}

\usepackage[breaklinks=true,colorlinks=true,linkcolor=blue,urlcolor=blue,citecolor=blue]{hyperref}

\begin{document}

\title[An embedding of the Bannai--Ito algebra in $\mathcal{U}(\mathfrak{osp}(1,2))$ and $-1$ polynomials]{An embedding of the Bannai--Ito algebra in $\mathcal{U}(\mathfrak{osp}(1,2))$ and $-1$ polynomials}

\author{Pascal Baseilhac}
\address{Laboratoire de Math\'ematiques et Physique Th\'eorique  CNRS/UMR 7350, F\'ed\'eration Denis Poisson FR2964 , Universit\'e de Tours, 37200 Tours, France}
\ead{baseilha@lmpt.univ-tours.fr}

\author{Vincent X. Genest}
\address{Department of Mathematics, Massachusetts Institute of Technology, Cambridge, MA 02139, USA}
\ead{vxgenest@mit.edu}

\author{Luc Vinet}
\address{Centre de recherches math\'ematiques, Universit\'e de Montr\'eal, Montr\'eal, QC H3C 3J7, Canada}
\ead{luc.vinet@umontreal.ca}

\author{Alexei Zhedanov}
\address{Department of Mathematics, School of Information, Renmin University of China, Beijing 100872,
	China}
\ead{zhedanov@yahoo.com}

\begin{abstract}
An embedding of the Bannai--Ito algebra in the universal enveloping algebra of $\mathfrak{osp}(1,2)$ is provided. A connection with the characterization of the little $-1$ Jacobi polynomials is found in the holomorphic realization of $\mathfrak{osp}(1,2)$. An integral expression for the Bannai--Ito polynomials is derived as a corollary. 
\end{abstract}


\section{Introduction}
This paper exhibits a direct connection between the superalgebra $\mathfrak{osp}(1,2)$ and the Bannai--Ito algebra. It also offers an interpretation of the little $-1$ Jacobi polynomials in this context and an integral formula for the Bannai--Ito polynomials. 

The Bannai--Ito polynomials were identified in the classification \cite{Bannai1984} of orthogonal polynomials with the Leonard duality property. They sit at the top of one the hierarchies of orthogonal polynomials that can be obtained under a $q\rightarrow -1$ limit of the members of the Askey tableau and which are hence called $-1$ polynomials \cite{Genest2013, Tsujimoto2012}. The Bannai--Ito algebra \cite{Tsujimoto2012} is a unital associative algebra with three generators that encodes the bispectral properties of the polynomials with the same name.

Since its introduction, the Bannai--Ito algebra has appeared in a number of contexts and some of its ties to $\mathfrak{osp}(1,2)$ have been uncovered. This algebra is in fact the symmetry algebra of a superintegrable model with reflections on the 2-sphere \cite{Genest2014d} and of the Dirac--Dunkl equation in three dimensions \cite{DeBie2016b}; it appears in Dunkl harmonic analysis on $S^2$ \cite{Genest2015a} and is isomorphic to the degenerate $(C_1^{\vee}, C_1)$ double affine Hecke algebra \cite{Genest2016a}. The Bannai--Ito algebra also arises in the Racah problem for $\mathfrak{osp}(1,2)$. Indeed, its central extension is the centralizer of the coproduct embedding of $\mathfrak{osp}(1,2)$ in the three-fold direct product $\mathfrak{osp}(1,2)^{\otimes 3}$ of this algebra with itself, with the intermediate Casimir operators acting as the generators \cite{Genest2014c}. A different relation between the two algebras (the Bannai--Ito one and $\mathfrak{osp}(1,2)$) will be presented in the following.

The $-1$ little Jacobi polynomials are orthogonal on $[-1,1]$ and depend on two parameters \cite{Vinet2011}. They are obtained as a $q\rightarrow -1$ limit of the little $q$-Jacobi polynomials and are eigenfunctions of a first order differential-difference operator of Dunkl type.

The Bannai-Ito algebra can be obtained by taking $q\rightarrow -1$ in the Askey--Wilson algebra $AW(3)$, which describes the bispectral properties of the Askey--Wilson polynomials \cite{Zhedanov1991}. It is known that $AW(3)$ is identified with a fixed point subalgebra of $\mathcal{U}_q(sl(2))$  under a certain automorphism. \cite{Granovskii1993a}. In a similar spirit, the goal here is to offer an embedding of the Bannai--Ito algebra in $\mathcal{U}(\mathfrak{osp}(1,2))$, the universal envelopping algebra of $\mathfrak{osp}(1,2)$. This will then be exploited in the context of the holomorphic realization. It will be found that in this realization one generator of the Bannai-Ito algebra coincides with the differential-difference operator that is diagonalized by the little $-1$ Jacobi polynomials. It will moreover be seen that a second generator results from the tridiagonalization \cite{Genest2016b} of the former. This will allow to obtain the eigenfunctions of this second generator.

As is generally understood from the theory of Leonard pairs \cite{Terwilliger2001}, the connection coefficients between two finite-dimensional representation bases constructed as eigenfunctions of either one of the Bannai--{I}to generators satisfy the three-term recurrence relation of the Bannai--Ito polynomials. The model developed will hence allow to provide an integral formula for the (finite) Bannai--Ito polynomials.

The outline is as follows. The embedding of the Bannai--Ito algebra in $\mathcal{U}(\mathfrak{osp}(1,2))$ is given in Section 2. The holomorphic representation of $\mathfrak{osp}(1,2)$ is considered in Section 3 where the defining operator of the little $-1$ Jacobi polynomials and its tridiagonalization will be seen to realize the Bannai--Ito generators. The integral formula for the Bannai--Ito polynomials is obtained in Section 4. Section 5 comprises concluding remarks. 

\section{Embedding the Bannai--Ito algebra in $\mathcal{U}(\mathfrak{osp}(1,2))$}
In this section, we present the formal embedding of the Bannai--Ito algebra in $\mathcal{U}(\mathfrak{osp}(1,2))$. The $\mathfrak{osp}(1,2)$ superalgebra is generated by the elements $A_{0}$, $A_{\pm}$ subject to the relations
\begin{align*}
[A_0, A_{\pm}] = \pm A_{\pm}, \qquad \{A_{+}, A_{-}\} = 2 A_0,
\end{align*}
where $[x,y] = xy - yx$ and $\{x,y\} = xy + yx$ stand for the commutator and the anticommutator, respectively. We introduce the grade involution operator $P$ satisfying
\begin{align*}
[A_0, P] = 0,\qquad \{A_{\pm}, P\} = 0, \qquad P^2 = 1.
\end{align*}
The above relations serve to indicate that $A_{\pm}$ are odd generators and that $A_0$ is an even generator. The Casimir operator $Q$ defined as
\begin{align}
\label{Casimir-Element}
Q = \frac{1}{2}([A_{-},A_{+}] - 1)P = (A_0 - A_{+} A_{-} - 1/2)P,
\end{align} 
commutes with $A_0$, $A_{\pm}$ and $P$, and generates the center of $\mathcal{U}(\mathfrak{osp}(1,2))$. 

Let $\mu_2$, $\mu_3$ and $\mu_4$ be real numbers and consider the operators $K_1, K_2, K_3 \in \mathcal{U}(\mathfrak{osp}(1,2))$ defined by the following expressions:
\begin{align}
\label{Realization}
\begin{aligned}
K_1 &= A_{+}A_0  -\mu_4 A_{+} P + (\mu_2 + \mu_3 + 1/2) A_{+} - A_{-}  + (\mu_4 - Q) P - 1/2,
\\
K_2 &= -A_{+}A_0P -(\mu_2 + \mu_3 + 1/2)  A_{+}P + A_0P +\mu_4 A_{+} + \mu_3 P,
\\
K_3 &= A_0 P- A_{-}P + \mu_2 P,
\end{aligned}
\end{align}
where $Q$ is the $\mathfrak{osp}(1,2)$ Casimir element \eqref{Casimir-Element}. The operators \eqref{Realization} obey the defining relations of the Bannai--Ito algebra. Indeed, a direct calculation shows that
\begin{align}
\label{BI}
\begin{aligned}
\{K_1, K_2\} = K_3 + \omega_3,\qquad  \{K_2, K_3\} = K_1 + \omega_1,\qquad \{K_3, K_1\} = K_2 + \omega_2, 
\end{aligned}
\end{align}
where $\omega_1$, $\omega_2$, $\omega_3$ are the central elements with expressions
\begin{align}
\label{struc}
\omega_1 = 2(\mu_4 Q + \mu_2 \mu_3 ),\quad \omega_2 = 2(\mu_3 Q + \mu_2 \mu_4),\qquad 
\omega_3 = 2(\mu_2  Q+ \mu_3 \mu_4).
\end{align}
It is verified that in the realization \eqref{Realization}, the Casimir operator $C$ of the Bannai--Ito algebra, which reads
\begin{align}
C = K_1^2 + K_2^2 +K_3^2,
\end{align} 
can be expressed as
\begin{align}
C = Q^2 + \mu_2^2 + \mu_3^2 + \mu_4^2 -1/4.
\end{align}
The combinations \eqref{Realization} thus provide a formal embedding of the Bannai--Ito algebra \eqref{BI} in the universal envelopping algebra of $\mathfrak{osp}(1,2)$. Since the structure constants $\omega_1, \omega_2, \omega_3$ in \eqref{struc} depend on the Casimir operator $Q$ of $\mathfrak{osp}(1,2)$, it follows that \eqref{BI} is in fact a central extension of the Bannai-Ito algebra, where the central operator is $Q$.
\section{Holomorphic realization and little $-1$ Jacobi polynomials}
In this section, we establish the connection between the embedding of the Bannai-Ito algebra in $\mathcal{U}(\mathfrak{osp}(1,2))$ detailed in the previous section and the little $-1$ Jacobi polynomials using the holomorphic realization of $\mathfrak{osp}(1,2)$. We also discuss the relationship with tridiagonalization.

\subsection{The holomorphic realization of $\mathfrak{osp}(1,2)$}
In the holomorphic realization of $\mathfrak{osp}(1,2)$, the generators $A_0$, $A_{\pm}$ and the grade involution are given by
\begin{align}
\label{osp-holo}
\begin{aligned}
A_0(x) = x\partial_{x} + (\mu_1 +1/2),\qquad A_{+}(x) = x,\qquad A_{-}(x) = D_x^{(\mu_1)}, \qquad P(x) = R_x, 
\end{aligned}
\end{align}
where $R_x f(x) = f(-x)$ is the reflection operator and where $D_x^{(\mu)}$ is the $A_1$ Dunkl operator
\begin{align}
D_x^{(\mu)} = \partial_{x} + \frac{\mu}{x}(1-R_x).
\end{align}
In the realization \eqref{osp-holo}, the Casimir operator \eqref{Casimir-Element} acts as a multiple of the identity; more specifically
\begin{align}
Q f(x) = \mu_1 f(x).
\end{align}
A natural basis for the irreducible representation underlying \eqref{osp-holo} is provided by the monomials. Upon defining $e_n(x) = x^{n}$, where $n$ is a non-negative integer, one has
\begin{align}
\label{actions}
\begin{aligned}
&A_0(x) e_{n}(x) = (n+\mu_1 +1/2)e_{n} (x),\qquad A_{+}(x) e_{n}(x) = e_{n+1}(x),
\\
&P(x) e_{n}(x) = (-1)^{n} e_{n}(x),\qquad A_{-}(x) e_{n}(x) = [n]_{\mu_1} e_{n-1}(x),
\end{aligned}
\end{align}
where 
\begin{equation*}
[n]_{\mu} = n + 2\mu(1-(-1)^{n}),
\end{equation*}
are the $\mu$-numbers. 
\subsection{Differential-Difference realization of Bannai--Ito generators}
In light of the embedding \eqref{Realization} of the Bannai--Ito algebra in $\mathcal{U}(\mathfrak{osp}(1,2))$, the holomorphic realization \eqref{osp-holo} and the basis $e_{n}(z)$ allow us to present an infinite-dimensional representation of the Bannai--Ito algebra in which the generators are realized as differential-difference operators. Let us denote by $K_1(x), K_2(x), K_3(x)$ the operators obtained by combining \eqref{Realization} with \eqref{osp-holo}. One has
\begin{align}
\label{Explicit}
\begin{aligned}
&K_1(x) = (x^2-1)\partial_{x} + x(\mu_1 + \mu_2 + \mu_3 + 1) - \frac{\mu_1}{x}(1-R_x) - (\mu_1 + \mu_4(x-1)) R_x - 1/2,
\\
&K_2 (x) = x(1-x)\partial_{x}R_x - x(\mu_1 + \mu_2 + \mu_3 +1) R_x + x\mu_4 + (\mu_1 + \mu_3 + 1/2) R_x,
\\
& K_3(x) = (x-1)\partial_{x}R_x + \left(\frac{\mu_1}{x} - (\mu_1+\mu_2 +1/2)\right) (1-R_x) + (\mu_1 + \mu_2 +1/2).
\end{aligned}
\end{align}
From \eqref{Explicit}, one finds that $K_3(x)$ has the action
\begin{align}
\label{Action-3}
K_3(x) e_{n}(x) = \lambda_{n} e_{n}(x) + \nu_n e_{n-1}(x),
\end{align}
where $ \lambda_{n}$ and $\nu_n$ are given by
\begin{align}
\label{Action-3-b}
\lambda_n = (-1)^{n}(n+\mu_1 + \mu_2 +1/2),\qquad 
\nu_n = (-1)^{n+1}[n]_{\mu_1}.
\end{align}
Similarly, $K_2(x)$ is seen to act bidiagonally as follows
\begin{align}
\label{Action-4}
K_2(x) e_{n}(x) = \kappa_{n+1} e_{n+1}(x) + \sigma_{n} e_{n}(x), 
\end{align}
where 
\begin{align}
\label{Action-4-b}
\kappa_n = (-1)^{n}(n + \mu_1 + \mu_2 + \mu_3 + (-1)^{n} \mu_4), \qquad \sigma_{n} = (-1)^{n}(n + \mu_1 + \mu_3 +1/2).
\end{align}
The third generator $K_1(x)$ acts in a three-diagonal fashion; one has
\begin{align}
\label{Action-5}
K_1(x)e_n(x) = \upsilon_{n+1} e_{n+1}(x) + \rho_n e_{n}(x)- \iota_n e_{n-1}(x)
\end{align} 
where the coefficients are given by
\begin{align}
\label{Action-5-b}
\upsilon_n = n + \mu_1 + \mu_2 + \mu_3 + (-1)^{n}\mu_4,\qquad \rho_n = (-1)^{n}(\mu_4-\mu_1)-1/2, \qquad \iota_n = -[n]_{\mu_1}.
\end{align}
We now proceed to construct the bases in which $K_2(x)$ and $K_3(x)$ are diagonal.

\subsection{The $K_3$ eigenbasis}
The eigenfunctions of the operator $K_3(x)$ can be constructed straightforwardly by solving the two-term recurrence relation that stems from the action \eqref{Action-3} of $K_3(x)$ on the monomial basis. However, it can be seen that $K_3(x)$ directly corresponds to the operator known to be diagonalized by the little $-1$ Jacobi polynomials $J_n^{(\alpha,\beta)}(x)$ \cite{Vinet2011}. The (monic) little $-1$ Jacobi polynomials are defined be the three-term recurrence relation
\begin{align}
\label{Rec-Relation}
J_{n+1}^{(\alpha,\beta)}(x) + b_n J_{n}^{(\alpha,\beta)}(x) + u_n J_{n-1}^{(\alpha,\beta)}(x) = x J_{n}^{(\alpha,\beta)}(x),
\end{align}
with $J_{-1}^{(\alpha,\beta)}(x) = 0$ and $J_{0}^{(\alpha,\beta)}(x) = 1$ and where $b_n$ and $u_n$ are given by
\begin{align}
\label{Recurrence-Coeff}
u_n = 
\begin{cases}
\frac{n(n+\alpha+\beta)}{(2n+\alpha+\beta)^2} & \text{$n$ even}
\\
\frac{(n+\alpha)(n+\beta)}{(2n+\alpha+\beta)^2} & \text{$n$ odd}
\end{cases}
,\qquad b_n=(-1)^n \frac{(2n+1)\alpha + \alpha \beta + \alpha^2 + (-1)^n\beta}{(2n+\alpha+\beta)(2n+\alpha+\beta+2)}.
\end{align}
In \cite{Vinet2011}, it was shown that the little $-1$ Jacobi polynomials satisfy the eigenvalue equation 
\begin{align*}
L J_{n}^{(\alpha,\beta)}(x) = t_n J_{n}^{(\alpha,\beta)}(x),\qquad 
t_n = \begin{cases}
-2n & \text{$n$ is even}
\\
2(\alpha+\beta+n +1) & \text{$n$ is odd}
\end{cases},
\end{align*}
where $L$ is the differential-difference operator 
\begin{align}
\label{L}
L = 2(1-x)\partial_x R_x + (\alpha +\beta + 1 -\alpha x^{-1}) (1-R_x).
\end{align}
Upon comparing $K_3(x)$ given by \eqref{Explicit} with \eqref{L}, one observes that $K_3(x)$ is diagonalized by the little $-1$ Jacobi polynomials with parameters $\alpha = 2\mu_1$ and $\beta = 2\mu_2$. Upon defining
\begin{align}
\label{Psi}
\psi_n(x) = J_{n}^{(2\mu_1, 2\mu_2)}(x),
\end{align}
one has the following eigenvalue relation
\begin{align}
\label{Psi-Eigen}
K_3(x) \psi_n(x) = \lambda_n \psi_n(x),
\end{align}
where the eigenvalues $\lambda_n$ are given by \eqref{Action-3-b}. It can easily be seen that $K_2(x)$ acts in a tridiagonal fashion on the little $-1$ Jacobi basis. Indeed, upon denoting by $X$ the ``multiplication by $x$'' operator, a straightforward calculation shows that $K_2(x)$ can be expressed as
\begin{align}
\label{Tri-diag}
K_2(x) = \tau_1 X K_3(x) + \tau_2 K_3(x) X + \tau_3 X + \tau_0,
\end{align}
where 
\begin{align*}
\tau_0 = -2\mu_1 \mu_3,\qquad \tau_1 = \mu_3 - 1/2, \qquad \tau_2 = \mu_3 +1/2, \qquad \tau_3 = \mu_4.
\end{align*}
Since the $X$ operator acts in a three-diagonal fashion on $J_n^{(\alpha,\beta)}(x)$ in accordance to the recurrence relation \eqref{Rec-Relation}, one has
\begin{multline}
\label{K2-Action}
K_2(x) \psi_n(x) = (\tau_1 \lambda_n + \tau_2 \lambda_{n+1} + \tau_3) \psi_{n+1}(x)
\\
+((\tau_1 +\tau_2)\lambda_n b_n + \tau_3 b_n + \tau_0)\psi_n(x)  +
(\tau_1 \lambda_n + \tau_2 \lambda_{n-1} + \tau_3) u_n \psi_{n-1}(x),
\end{multline}
where the coefficients $u_n$ and $b_n$ are given by \eqref{Recurrence-Coeff} with $\alpha = 2\mu_1$ and $\beta = 2\mu_2$.

\begin{Remark}
The expression \eqref{Tri-diag} indicates that $K_2(x)$ can be obtained from $K_3(x)$ via the tridiagonalization procedure. This procedure has been discussed in \cite{Genest2016b, Ismail2012}; it here allows to straightforwardly construct the representation of the Bannai--Ito algebra in the basis provided by the little $-1$ Jacobi polynomials.
\end{Remark}
\subsection{The $K_2$ eigenbasis}
We now determine the eigenbasis associated to $K_2(x)$. We first observe that, in parallel with \eqref{Tri-diag}, $K_3(x)$ can be expressed in terms of $K_2(x)$ as follows:
\begin{align}
\label{Tridiag-2}
K_3(x) = \beta_1 X^{-1} K_2(x) + \beta_2 K_2(x) X^{-1} + \beta_3 X^{-1} + \beta_0,
\end{align}
where $X^{-1}$ is the ``multiplication by $1/x$'' operator and where
\begin{align*}
\beta_0 = -2\mu_3 \mu_4, \qquad \beta_1 = \mu_3-1/2,\qquad \beta_2 = \mu_3 +1/2, \qquad \beta_3 = \mu_1.
\end{align*}
In view of \eqref{Tridiag-2}, we consider the change of variable $y = 1/x$. Under this transformation, $K_2(x)$ takes the form
\begin{align*}
K_2(y) = (1-y) \partial_{y}R_y + \left(\mu_1 + \mu_3 + 1/2 - \frac{\mu_1+\mu_2+\mu_3 +1}{y}\right)R_y + \frac{\mu_4}{y}.
\end{align*} 
Consider the gauge factor $\phi_{\epsilon}(y)$ defined as
\begin{align*}
\phi_{\epsilon}(y) = y^{\epsilon} |y|^{\mu_1 + \mu_2 + \mu_3 - (-1)^{\epsilon} \mu_4 + 1 - \epsilon},
\end{align*} 
where $\epsilon = 0$ or $1$. One has 
\begin{multline}
\label{Gauge-Trans}
\phi_{\epsilon}(y)^{-1} K_2(y) \phi_{\epsilon}(y) =(-1)^{\epsilon} \Bigg[(1-y)\partial_{y}R_y
\\
 + \left(\frac{(-1)^{\epsilon}\mu_4}{y} - ((-1)^{\epsilon}\mu_4-\mu_2 - 1/2)\right)(1-R_y) + ((-1)^{\epsilon}\mu_4-\mu_2 -1/2) \Bigg].
\end{multline}
It is seen that \eqref{Gauge-Trans} has the same form as $K_3(x)$. It follows that the eigenfunctions of $K_2(x)$ have the expression
\begin{align}
\label{Chi}
\chi_n(x) = \phi_{\epsilon}(1/x)\, J_{n}^{(-2(-1)^{\epsilon}\mu_4, 2\mu_2)}(1/x),
\end{align}
and that the eigenvalue equation reads
\begin{align}
\label{Chi-Eigen}
K_2(x)\chi_n(x) = \Omega_n \chi_n(x),
\end{align}
with the eigenvalues $\Omega_n$ given by
\begin{align}
\label{Omega}
\Omega_n = (-1)^{\epsilon}(-1)^{n+1}(n + \mu_2 + (-1)^{\epsilon+1}\mu_4 + 1/2).
\end{align}
\subsection{Finite-dimensional reduction}
As is clear from the action \eqref{Action-4} of $K_2(x)$ on the monomial basis, the action of the Bannai--Ito generators \eqref{Explicit} does not preserve the space of polynomials of a given degree. A finite-dimensional representation can however be obtained by imposing the appropriate truncation condition on the parameters. Indeed, it is easily seen that if one takes
\begin{align}
\label{Truncation}
\mu_4 \rightarrow \mu_N = (-1)^{N}(N+\mu_1 + \mu_2 + \mu_3 +1),\qquad N = 1, 2,\ldots
\end{align}
then, the action of the Bannai--Ito generators preserves the $(N+1)$-dimensional vector space spanned by the monomials $\{e_0(x),\ldots, e_{N}(x)\}$. Upon imposing the truncation condition \eqref{Truncation}, the formula \eqref{Psi} for the eigenfunctions $\psi_n(x)$ of $K_3(x)$ as well as the eigenvalue equation \eqref{Psi-Eigen} remain valid. For the eigenfunctions $\chi_n(x)$ of $K_2(x)$ obtained in \eqref{Chi}, one must take $\epsilon = 0$ in \eqref{Chi}  and \eqref{Chi-Eigen} when $N$ is even, while taking $\epsilon = 1$ in \eqref{Chi}  and \eqref{Chi-Eigen} when $N$ is odd. These choices guarantee that $\chi_n(x)$ is a polynomial of degree less or equal to $N$. It is observed that when the truncation condition \eqref{Truncation} is satisfied, $\chi_n(x)$ is in fact of the form $\chi_n(x) = c_0 x^{N} + c_1 x^{N-1} + \cdots + c_n x^{N-n}$, as expected from the lower-triangular shape of $K_2(x)$. We shall assume that \eqref{Truncation} holds from now on.
\section{Bannai-Ito polynomials}
In this Section, the Bannai--Ito polynomials are shown to arise as the interbasis expansion coefficients between the eigenbases of $K_2(x)$ and $K_3(x)$. This leads to an integral expression for the Bannai--Ito polynomials involving the little $-1$ Jacobi polynomials.
\subsection{A scalar product}
Let $P(x)$ and $Q(x)$ be real polynomials in $x$; we introduce their scalar product denoted by $\langle P(x), Q(x)\rangle$ and defined as
\begin{align*}
\langle P(x), Q(x) \rangle  = \int_{-1}^{1} \omega(x) P(x) Q(x)\,\mathrm{d}x,
\end{align*}
where $\omega(x)$ is given by
\begin{align*}
\omega(x) =|x|^{2\mu_1}(1-x^2)^{\mu_2-1/2}(1+x).
\end{align*}
Under this scalar product, the elements of the eigenbasis of $K_3(x)$, given by the little $-1$ Jacobi polynomials, are orthogonal \cite{Vinet2011}. Indeed, one has
\begin{align}
\label{Ortho-Psi}
\langle \psi_{n}(x), \psi_m(x)\rangle = h_{n}\delta_{nm},
\end{align}
where the normalization coefficients $h_n$ have the expression
\begin{multline}
\label{Norm}
h_n = \frac{\Gamma(\mu_1 +1/2)\Gamma(\mu_2 +1/2)}{\Gamma(\mu_1+\mu_2 +1)}
\\ 
\times
\frac{(\lfloor n/2\rfloor)! (\mu_1 +1/2)_{\lceil n/2 \rceil} (\mu_2 +1/2)_{\lceil n/2 \rceil} (\mu_1 +\mu_2 +1)_{\lfloor n/2 \rfloor}}{(\mu_1+\mu_2+1/2)^2_n},
\end{multline}
where $(a)_n$ stands for the Pochhammer symbol. 
\subsection{Interbasis expansion coefficients}
We now consider the interbasis expansion coefficients between the eigenbases of $K_3(x)$ and $K_2(x)$. These coefficients, which shall be denoted by $\mathcal{B}_n(k)$, are defined by the following expansion of the $K_2(x)$ eigenfunction in a series of little $-1$ Jacobi polynomials
\begin{align}
\label{Overlap-Def}
\chi_k(x) = \sum_{n=0}^{N} \mathcal{B}_{n}(k) \; \psi_{n}(x),
\end{align}
In light of the orthogonality relation \eqref{Ortho-Psi} satisfied by $\psi_n(x)$, one can write
\begin{align}
\label{Overlap-Integral}
\mathcal{B}_{n}(k) = \frac{1}{h_n}\langle \chi_k(x), \psi_n(x)\rangle = \frac{1}{h_n}\int_{-1}^{1} \omega(x)\,\chi_k(x)\psi_n(x)\,\mathrm{dx}.
\end{align}
We recall that the coefficients $\mathcal{B}_n(k)$ also depend on the three parameters $\mu_1$, $\mu_2$, $\mu_3$, as well as on $\mu_4 = \mu_N$. In light of the truncation condition \eqref{Truncation}, the integral in \eqref{Overlap-Integral} is always well-defined provided that $\mu_i \geq 0$ for $i=1,2,3$. 

It is clear that the coefficients $\mathcal{B}_n(k)$ satisfy a three-term recurrence relation. Indeed, upon applying $K_2(x)$ on \eqref{Overlap-Def}, using the eigenvalue equation \eqref{Chi-Eigen} and the action \eqref{K2-Action} of $K_2(x)$ on $\psi_n(x)$ and finally exploiting the linear independence of the little $-1$ Jacobi polynomials, one finds that $\mathcal{B}_n(k)$ obey
\begin{align}
\label{Recurrence}
\Omega_k \mathcal{B}_n(k) = E_n^{(1)}\mathcal{B}_{n+1}(k)
+E_n^{(2)} \mathcal{B}_{n}(k) + E_n^{(3)}\mathcal{B}_{n-1}(k),
\end{align}
where the recurrence coefficients are given by
\begin{align}
\label{Coefs}
\begin{aligned}
&E_n^{(1)} = (\tau_1 \lambda_{n+1} + \tau_2 \lambda_n + \tau_3) u_{n+1},
\qquad
E_n^{(2)} = ((\tau_1+\tau_2)\lambda_n b_n + \tau_3 b_n + \tau_0),
\\
&
E_n^{(3)} = (\tau_1 \lambda_{n-1} + \tau_2 \lambda_{n} +\tau_3).
\end{aligned}
\end{align}
One can write $\mathcal{B}_{n}(k) = \mathcal{B}_{0}(k) P_n(\Omega_k)$ with $P_0(\Omega_k) =1$ and
\begin{align}
\label{B0}
\mathcal{B}_{0}(k) =  \frac{1}{h_0}\int_{-1}^{1} \omega(x)\,\chi_k(x)\psi_0(x)\,\mathrm{dx}.
\end{align}
It is clear from \eqref{Recurrence} that $P_n(\Omega_k)$ are polynomials of degree $n$ in $\Omega_k$. We introduce the normalized polynomials $\widehat{P}_n(\Omega_k) = E_{0}^{(1)}\cdots E_{n-1}^{(1)} P_n(\Omega_k) $, which satisfy the normalized recurrence relation
\begin{align}
\label{Normalized-Recurrence}
\Omega_k \widehat{P}_n(\Omega_k) = \widehat{P}_{n+1}(\Omega_k) + r_n \widehat{P}_{n+1}(\Omega_k) + U_n \widehat{P}_{n-1}(\Omega_k),
\end{align}
with coefficients 
\begin{align*}
U_n = E_{n}^{(3)}E_{n-1}^{(1)},\qquad r_n = E_{n}^{(2)}.
\end{align*}
A direct calculation shows that the coefficients $U_n$ and $r_n$ can be expressed as follows
\begin{align*}
U_n = A_{n-1} C_{n},\qquad r_n = \mu_1+\mu_3 + 1/2 -A_{n} - C_{n}
\end{align*}
where $A_{n}$ and $C_n$ are given by
\begin{align}
\label{BI-Coefs}
\begin{aligned}
&A_n =
\begin{cases}
\frac{(n+2\mu_1+1)(n+\mu_1+\mu_2+\mu_3-\mu_N + 1)}{2(n+\mu_1 +\mu_2+1)} & \text{$n$ even}
\\
\frac{(n+2\mu_1+2\mu_2 +1)(n+\mu_1 +\mu_2 +\mu_3 +\mu_N+1)}{2(n+\mu_1+\mu_2)} & \text{$n$ odd}
\end{cases},
\\
&C_n =
\begin{cases}
-\frac{n(n+\mu_1 +\mu_2 -\mu_3 -\mu_N)}{2(n+\mu_1 +\mu_2)} & \text{$n$ even}
\\
-\frac{(n+2\mu_2)(n+\mu_1 +\mu_2 -\mu_3 + \mu_N)}{2(n+\mu_1+\mu_2)} & \text{$n$ odd}
\end{cases}.
\end{aligned}
\end{align}
The coefficients \eqref{BI-Coefs} correspond to those of the monic Bannai--Ito polynomials in the parametrization associated to the Racah problem (with the permutation $\mu_1 \leftrightarrow \mu_2$); see \cite{Genest2014c} for details on how to relate the present parametrization to the parametrization given in \cite{Tsujimoto2012}. It follows that the monic Bannai--Ito polynomials admit the integral expression
\begin{align}
\label{Integral-Formula}
\widehat{P}_n(\Omega_k) = \frac{E_0^{(1)}\cdots E_{n-1}^{(1)}}{h_n \mathcal{B}_0(k)} \int_{-1}^{1} \omega(x)\; \chi_k(x)\psi_n(x) \,\mathrm{d}x,
\end{align}
where $E_{n}^{(1)}$ is given by \eqref{Coefs}, $h_n$ by \eqref{Norm}, $\mathcal{B}_0(k)$ by \eqref{B0}, $\chi_k(x)$ and $\Omega_k$ by \eqref{Chi} and \eqref{Omega}, and where $\psi_n(x)$ is given by \eqref{Psi}. In essence, \eqref{Integral-Formula} gives an expression for the Bannai--Ito polynomials as an integral over the product of two little $-1$ Jacobi polynomials. This is an analog of Koornwinder's integral representation of the Wilson polynomials \cite{Koornwinder1984}.
\section{Conclusion}
In this paper, we have exhibited a direct connection between the Bannai--Ito algebra and the superalgebra $\mathfrak{osp}(1,2)$. We have provided an explicit embedding of the Bannai--Ito algebra in $\mathcal{U}(\mathfrak{osp}(1,2))$, and offered a new characterization of the little $-1$ Jacobi polynomials in the context of the holomorphic realization. We also highlighted connections with the tridiagonalization approach to orthogonal polynomials. Finally, we have given a new integral representation of the Bannai--Ito polynomials.

As already mentioned in the introduction, the Bannai--Ito algebra has also been seen to arise as the algebra formed by the intermediate Casimir operators in the addition of three $\mathfrak{osp}(1,2)$ superalgebras \cite{Genest2014c}. It would be of interest to see if an explicit correspondence relating this connection between the Bannai--Ito algebra and $\mathfrak{osp}(1,2)$ and the one identified here could be established in parallel to what was found in the case of the Racah algebra \cite{Genest2014e}. Besides, it is known that the Askey--Wilson algebra can be viewed as a homomorphic image of the $q$-Onsager algebra and that the Bannai--Ito algebra is obtained from the former when $q=-1$. This suggests that the embedding in $\mathcal{U}(\mathfrak{osp}(1,2))$ could be viewed as a subalgebra of a $q$-Onsager algebra for $q$ a root of unity, see for instance \cite{Baseilhac2017}. This would certainly be worth exploring. We plan to look into these two questions in the future.
\section*{Acknowledgments}
PB, VXG and AZ acknowledge the hospitality of the CRM and LV that of the Universit\'e de Tours where parts of the reported research has been realized. PB is supported by C.N.R.S. VXG holds a postdoctoral fellowship from the Natural Science and Engineering Research Council (NSERC) of Canada. LV is grateful to NSERC for support through a discovery grant.
\section*{References}

\begin{thebibliography}{10}
	
	\bibitem{Bannai1984}
	E.~Bannai and T.~Ito.
	\newblock {\em Algebraic Combinatorics I: Association Schemes}.
	\newblock Benjamin \& Cummings, 1984.
	
	\bibitem{Baseilhac2017}
	P.~Baseilhac, A.~M. Gainutdinov, and T.~T. Vu.
	\newblock Cyclic tridiagonal pairs, higher order {O}nsager algebras and
	orthogonal polynomials.
	\newblock {\em Lin. Alg. \& Appl.}, 522:71--110, 2017.
	
	\bibitem{DeBie2016b}
	H.~De~Bie, V.~X. Genest, and L.~Vinet.
	\newblock A {D}irac-{D}unkl equation on {$S^2$} and the {B}annai-{I}to algebra.
	\newblock {\em Commun. Math. Phys.}, 344:447--464, 2016.
	
	\bibitem{Genest2016b}
	V.~X. Genest, M.~Ismail, L.~Vinet, and A.~Zhedanov.
	\newblock Tridiagonalization of the hypergeometric operator and the
	{R}acah-{W}ilson algebra.
	\newblock {\em Proc. Amer. Math. Soc.}, 144:4441--4454, 2016.
	
	\bibitem{Genest2013}
	V.~X. Genest, L.~Vinet, and A.~Zhedanov.
	\newblock Bispectrality of the complementary {B}annai-{I}to polynomials.
	\newblock {\em SIGMA Symmetry Integrability Geom. Methods Appl.}, 9:18--37,
	2013.
	
	\bibitem{Genest2014c}
	V.~X. Genest, L.~Vinet, and A.~Zhedanov.
	\newblock The {B}annai--{I}to polynomials as {R}acah coefficients of the
	$sl_{-1}(2)$ algebra.
	\newblock {\em Proc. Amer. Math. Soc.}, 142:1545--1560, 2014.
	
	\bibitem{Genest2014d}
	V.~X. Genest, L.~Vinet, and A.~Zhedanov.
	\newblock The {B}annai-{I}to algebra and a superintegrable system with
	reflections on the two-sphere.
	\newblock {\em J. Phys. A: Math. Theor.}, 47:205202, 2014.
	
	\bibitem{Genest2014e}
	V.~X. Genest, L.~Vinet, and A.~Zhedanov.
	\newblock The equitable racah algebra from three {$\mathfrak{su}(1,1)$}
	algebras.
	\newblock {\em J. Phys. A: Math. Theor.}, 47:025203, 2014.
	
	\bibitem{Genest2015a}
	V.~X. Genest, L.~Vinet, and A.~Zhedanov.
	\newblock A {L}aplace-{D}unkl equation on {$S^2$} and the {B}annai--{I}to
	algebra.
	\newblock {\em Commun. Math. Phys.}, 336:243--259, 2015.
	
	\bibitem{Genest2016a}
	V.~X. Genest, L.~Vinet, and A.~Zhedanov.
	\newblock The non-symmetric {W}ilson polynomials are the {B}annai--{I}to
	polynomials.
	\newblock {\em Proc. Amer. Math. Soc.}, 144:5217--5226, 2016.
	
	\bibitem{Granovskii1993a}
	Y.~I. Granovskii and A.~Zhedanov.
	\newblock Linear covariance algebra for {$SL_q(2)$}.
	\newblock {\em J. Phys. A: Math. Gen.}, 26:L357, 1993.
	
	\bibitem{Ismail2012}
	M.~Ismail and E.~Koelink.
	\newblock Spectral properties of operators using tridiagonalisation.
	\newblock {\em Anal. \& Appl.}, 10:327, 2012.
	
	\bibitem{Koornwinder1984}
	T.~H. Koornwinder.
	\newblock {\em Special orthogonal polynomial systems mapped onto each other by
		the {F}ourier-{J}acobi transform}, pages 174--183.
	\newblock Lecture Notes in Math. Springer, 1984.
	
	\bibitem{Terwilliger2001}
	P.~Terwilliger.
	\newblock Two linear transformations each tridiagonal with respect to an
	eigenbasis of the other.
	\newblock {\em Lin. Alg. \& Appl.}, 330:149--203, 2001.
	
	\bibitem{Tsujimoto2012}
	S.~Tsujimoto, L.~Vinet, and A.~Zhedanov.
	\newblock Dunkl shift operators and {B}annai-{I}to polynomials.
	\newblock {\em Adv. Math.}, 229:2123--2158, 2012.
	
	\bibitem{Vinet2011}
	L.~Vinet and A.~Zhedanov.
	\newblock A 'missing' family of classical orthogonal polynomials.
	\newblock {\em J. Phys. A: Math. Theor.}, 44:085201, 2011.
	
	\bibitem{Zhedanov1991}
	A.~S. Zhedanov.
	\newblock ``{H}idden symmetry'' of {A}skey--{W}ilson polynomials.
	\newblock {\em Theoretical and Mathematical Physics}, 89:1146--1157, 1991.
	
\end{thebibliography}

\end{document}